\pdfoutput=1
\documentclass[floatfix,nofootinbib,twocolumn,english,pra,aps]{revtex4}
\usepackage{silence}
\usepackage{mathtools,amsmath}
\usepackage{amssymb}
\usepackage{graphicx}
\usepackage{babel}
\makeatletter
\pdfpageheight\paperheight
\pdfpagewidth\paperwidth

\begin{document}
\title{\noindent Quantum amplification, relic gravitons and Landauer's conjecture}
\author{\noindent Massimo Giovannini}
\email{massimo.giovannini@cern.ch}
\affiliation{\noindent Department of Physics, CERN, 1211 Geneva 23, Switzerland and INFN, Section of Milan-Bicocca, 20126 Milan, Italy}
\begin{abstract}
\noindent According to the microscopic formulation of Landauer's principle, when information is deleted the Von Neumann entropy of the system gets reduced with a corresponding energy cost. Although within the same perspective the growth of the entropy should remain unconstrained we show that during quantum amplification the heat flow does restrict the increase of the Von Neumann entropy. 
When applied to the case of relic gravitons (with frequencies between the aHz region and the THz domain) the bounds obtained here set a limit on initial thermal gravitons and on the total duration of inflation. 
\end{abstract}
\maketitle
The Landauer's conjecture \cite{LAN1,BEN1,LAN2} stipulates that the (irreversible) erasure of one bit demands an energy input larger or equal to $T\ln2$, where $T$ denotes hereunder the temperature of the environment\footnote{The natural system of units $\hbar= c= k_{B} =1$ is employed throughout; in terms of the Newton's constant $G$ the Planck length $\ell_{P}$ and the Planck mass $M_{P}$ are defined accordingly, i.e. $\ell_{P} = \sqrt{8\pi}/M_{P}$ and $M_{P} = 1/\sqrt{G}$.}. From a microscopic perspective this conjecture underlies a physical principle applicable to all quantum systems where an originally pure state interacts with an initially thermal environment \cite{LAN3}. We want to examine here the microscopic description of parametric amplification \cite{LOU1,MG1,MAND} in the light of the quantum thermodynamical considerations usually associated with the Landauer's principle.  As an application, the restrictions on the entropy variation shall be explored in the case of the gravitons that are produced thanks to the early variation of the space-time curvature \cite{GR1,GR2,FP} especially in connection with the conventional inflationary paradigm \cite{AA7,AA8,AA9} (see also \cite{rr1} for a review). 

If the quantum mechanical Hamiltonians of the system and of the environment (denoted, respectively, by $\hat{H}_{a}$ and $\hat{H}_{b}$) are initially uncorrelated (and characterized by an Hermitian interaction $\hat{H}_{a b}$) the unitary evolution suggests that the heat transferred to the environment $\Delta Q_{b}$ should be related to the variation of the entropy of the system $\Delta S_{a}$ by the microscopic formulation of the Landauer's principle \cite{SAG} as:
\vspace{-0.1cm}
 \begin{equation}
 \Delta Q_{b} \, \geq \, - T\, \Delta \, S_{a},\quad \Delta Q_{b} = \mathrm{Tr}[ \hat{H}_{b}( \hat{\rho}_{b}^{(fin)}-\hat{\rho}_{b}^{(in)})],
 \label{EQA}
 \end{equation}
where $T_{b} = T$ denotes here the temperature of the thermal environment. The (uncorrelated) density matrix of the initial state $\hat{\rho}_{a b}^{(in)} = \hat{\rho}^{(in)}_{a} \otimes  \hat{\rho}^{(in)}_{b}$ evolves unitarily
\vspace{-0.1cm}
\begin{equation} 
 \hat{\rho}_{a b}^{(fin)} =  \hat{U}(\tau_{fin}, \tau_{in}) \, \hat{\rho}_{a b}^{(in)} \, \hat{U}^{\dagger}(\tau_{fin}, \tau_{in}),
 \label{EQAU}
 \vspace{-0.1cm}
 \end{equation}
since $\hat{U}^{\dagger}=\hat{U}^{-1}$. The density operator of the final state can be traced over the degrees of freedom of the environment (i.e.  $\hat{\rho}_{a}^{(fin)} = \mathrm{Tr}_{b} [ \, \hat{\rho}_{a\, b}^{(fin)}]$) and of the system (i.e. $\hat{\rho}_{b}^{(fin)} = \mathrm{Tr}_{a} [ \, \hat{\rho}_{a\, b}^{(fin)}]$). With these standard quantum mechanical notations the variation of the (Von Neumann) entropy of the system appearing in Eq. (\ref{EQA}) reads $\Delta\, S_{a}= S[ \hat{\rho}_{a}^{(fin)}] - S[ \hat{\rho}_{a}^{(in)}]$ where, as usual, $S[\hat{\rho}] = - \mathrm{Tr}[\hat{\rho} \, \ln{\hat{\rho}}]$.
 
 When information is erased, the final entropy of the system decreases from an initially larger value so that $\Delta\, S_{a} < 0$: to delete {\em one bit} of information $\Delta\, S_{a} = -\ln{2}$ and Eq. (\ref{EQA}) demands that the energy cost for this minimal erasure is $\Delta Q_{b} \geq  T \ln{2}$ \cite{LAN1,BEN1,LAN2}. Although the potential saturation of the bound given by Eq. (\ref{EQA}) is under debate \cite{RIE}, the Landauer's conjecture has been experimentally verified in a number of different frameworks \cite{SHIZ,DILL,JUN}. Since the removal of information requires a decrease of the entropy of the system, when $\Delta S_{a} >0$ the condition imposed by Eq. (\ref{EQA}) does not seem restrictive: a physical quantity which is positive semi-definite (i.e. $\Delta Q_{b}\geq 0$) always exceeds a negative contribution (i.e. $- T \Delta S_{a}$). In other words there is an inevitable energy cost to delete information whereas the acquisition of the information remains practically unconstrained: this is ultimately the logic behind the proposed solution of the well known Maxwell's paradox \cite{BEN1,LAN2,LAN3}. We now point out that when the system and the environment are initially uncorrelated the quantum parametric amplification implies $\Delta S_{a} \geq 0$ so that Eq. (\ref{EQA}) does not suggest any relevant restriction. The purpose of the present investigation is however to demonstrate that the growth of the entropy of the system associated with the parametric amplification complies with the following physical bound
\vspace{-0.1cm}
\begin{equation}
\Delta S_{a} < \Delta Q_{b}/T, \quad \overline{n}^{(q)} \gg 1,
\label{EQA1}
\vspace{-0.1cm}
\end{equation}
where $\overline{n}^{(q)}$ indicates throughout the average multiplicity of the produced 
quanta. In case  $\overline{n}^{(q)}= {\mathcal O}(1)$ the bound of Eq. (\ref{EQA1})
is still valid provided $\overline{n} \leq e^{\overline{n}^{(q)}-1}/\overline{n}^{(q)}$ (having 
denoted with $\overline{n}$ the average thermal multiplicity of the environment at temperature $T$).

To deduce the bound of Eq. (\ref{EQA1}) we now examine the situation 
where the system and the environment correspond, respectively, to a pair of quantum oscillators with frequencies $\omega_{a}$ and $\omega_{b}$:
\vspace{-0.1cm}
\begin{equation}
\hat{H}_{a} = \omega_{a} ( \hat{a}^{\dagger} \, \hat{a} + 1/2), \qquad \hat{H}_{b} = \omega_{b} (\hat{b}^{\dagger} \, \hat{b} + 1/2),
\label{EQB}
\end{equation}
with $[\hat{a},\hat{b}]=0$.
As required by Eq. (\ref{EQAU}), the density matrices of the two components are 
(initially) uncorrelated i.e. $\hat{\rho}_{a\,b}^{(in)} = \hat{\rho}^{(in)}_{a} \otimes \hat{\rho}^{(in)}_{b}$; in particular the system is in the vacuum (i.e.  $\hat{\rho}^{(in)}_{a} = |\,0 \rangle \langle 0\,|$) whereas the density matrix of the environment is a mixture with statistical weights provided by the Bose-Einstein (geometric) distribution:
\vspace{-0.3cm}
\begin{equation}
\hat{\rho}^{(in)}_{b} = \sum_{m= 0}^{\infty}\,\frac{\overline{n}^{m}}{(\overline{n} +1)^{m + 1}} \, | \, m\rangle \langle m\,|, 
\label{EQF}
\vspace{-0.2cm}
\end{equation}
where $\overline{n}$ denotes the averaged thermal multiplicity.  To have quantum amplification the interaction Hamiltonian $\hat{H}_{a b}$ must not commute with the sum of the number operators of the system and of the environment; its general form is \cite{LOU1,MG1,MAND}:
\vspace{-0.1cm}
\begin{equation}
\hat{H}_{a b} = \lambda \,\hat{a}^{\dagger} \, \hat{b}^{\dagger} \,\, e^{ - i \omega \, \tau} + \lambda^{\ast} \,\hat{a} \, \hat{b} \,\,e^{ i \omega \, \tau},
\label{EQG}
\vspace{-0.1cm}
\end{equation}
where $\lambda(\tau)= q(\tau)\, e^{i\, \theta_{in}}$ and  $ \omega = \omega_{a} + \omega_{b}$. The evolution of $\hat{a}$ and $\hat{b}$ follows then from the {\em total} Hamiltonian given by the sum of Eqs. (\ref{EQB}) and (\ref{EQG}), i.e. $\hat{H}= \hat{H}_{a}  + \hat{H}_{b} + \hat{H}_{a\,b}$. The solution of the corresponding Heisenberg equations $ \partial_{\tau} \hat{a} = i \, [ \hat{H},  \hat{a}]$ and $ \partial_{\tau} \hat{b} = i \, [ \hat{H},  \hat{b}]$ is:
\begin{eqnarray}
\hat{a}^{(fin)} &=& e^{-i \delta_{a}} \bigl[ \cosh{r} \,\hat{a}^{(in)} - e^{i \theta}\, \sinh{r} \, \hat{b}^{(in)\dagger} \bigr],
\nonumber\\
\hat{b}^{(fin)\dagger} &=& e^{i \delta_{b}} \bigl[ \cosh{r} \,\hat{b}^{(in)\dagger}  - e^{- i \theta}\, \sinh{r} \, \hat{a}^{(in)} \bigr],
\label{EQI}
\end{eqnarray}
where $\theta = (\pi/2 + \theta_{in})$ while $\delta_{a}=\omega_{a}\Delta\tau$ and $\delta_{b}= \omega_{b}\Delta\tau$ (we set here $\Delta \tau=(\tau_{fin}-\tau_{in})$); from Eq. (\ref{EQG}) it follows that $ r = \int_{\tau_{in}}^{\tau_{fin}} d\tau^{\prime} q(\tau^{\prime})$. Thanks to the Baker-Hausdorff lemma, Eq. (\ref{EQI}) may be reformulated in terms of two unitary operators \cite{MAND} given, respectively, by
$\hat{{\mathcal R}}(\delta_{a},\delta_{b})=\exp{[ - i \delta_{a} \,\hat{a}^{(in)\,\dagger} \hat{a}^{(in)}- i \delta_{b} \,\hat{b}^{(in)\,\dagger} \hat{b}^{(in)}]}$ and by $\hat{\Sigma}(z)=\exp{[z^{\ast}\, \hat{a}^{(in)} \, \hat{b}^{(in)} - z \, \hat{a}^{(in)\,\dagger} \, \hat{b}^{(in)\,\dagger}]}$ where $z= r\, e^{i\,\theta}$. The unitary transformation appearing in Eq. (\ref{EQI}) is then expressed as $\hat{a}^{(fin)} = \hat{\Sigma}^{\dagger}(z)\, \hat{{\mathcal R}}^{\dagger}(\delta_{a},\delta_{b}) \, \hat{a}^{(in)}\, \hat{{\mathcal R}}(\delta_{a},\delta_{b})\, \hat{\Sigma}(z)$ and  as $ \hat{b}^{(fin)\dagger} = \hat{\Sigma}^{\dagger}(z)\, \hat{{\mathcal R}}^{\dagger}(\delta_{a}, \delta_{b}) \, \hat{b}^{(in)\dagger}\, \hat{{\mathcal R}}(\delta_{a},\delta_{b})\, \hat{\Sigma}(z)$. 

The density matrix at late time can then be written in terms of  $\hat{{\mathcal R}}(\delta_{a},\delta_{b})$ and $\hat{\Sigma}(z)$:
\begin{equation}
\hat{\rho}^{(fin)}_{a b} = \hat{{\mathcal R}}(\delta_{a}, \delta_{b}) \, \hat{\Sigma}(z) \,\, \hat{\rho}_{a b}^{(in)} \,\,  \hat{\Sigma}^{\dagger}(z)  \hat{{\mathcal R}}^{\dagger}(\delta_{a}, \delta_{b}),
\label{EQM}
\end{equation}
while the reduced density operators $\hat{\rho}_{b}^{(fin)} = \mathrm{Tr}_{a}[ \hat{\rho}_{ab}^{(fin)}]$ and $\hat{\rho}_{a}^{(fin)} = \mathrm{Tr}_{b}[ \hat{\rho}_{ab}^{(fin)}]$ directly follow from  Eq. (\ref{EQM}); their explicit form becomes:
\vspace{-0.1cm}
\begin{eqnarray}
\hat{\rho}_{b}^{(fin)} &=&  \sum_{\ell=0}^{\infty} \sum_{m=0}^{\infty} p^{(b)}_{m\,\ell}[\overline{n}, \overline{n}^{(q)}] \, |\,\ell+m \rangle\langle m+ \ell\, |,
\label{EQN}\\
\hat{\rho}_{a}^{(fin)} &=& \sum_{\ell=0}^{\infty} 
p^{(a)}_{\ell}[\overline{n}, \overline{n}^{(q)}] \, |\,\ell \rangle\langle \ell\, |,
\label{EQO}
\end{eqnarray}
where, as in Eq. (\ref{EQA1}), $\overline{n}^{(q)} = \sinh^2{r}$ denotes the average multiplicity of the produced quanta. The statistical weights entering Eqs. (\ref{EQN})--(\ref{EQO}) are eventually given by:
\vspace{-0.1cm}
\begin{eqnarray}
&& p^{(b)}_{\ell\, m}[\overline{n}, \overline{n}^{(q)}]  = \binom{m+ \ell}{m} \frac{ \overline{n}^{m}\, \overline{n}^{(q)\,\,\ell}}{(\overline{n} +1)^{m +1} [\overline{n}^{(q)} +1]^{m + \ell +1}},
\nonumber\\
&& p^{(a)}_{\ell}[\overline{n}, \overline{n}^{(q)}] = \frac{\overline{n}^{(q)\, \ell}\, ( 1 + \overline{n})^{\ell}}{[ 1 + \overline{n}^{(q)}(\overline{n}+1)]^{\ell +1}}.
\label{EQP}
\end{eqnarray}
The derivation of Eqs. (\ref{EQN})--(\ref{EQP}) simplifies by appreciating that the operators $(\hat{a}^{\dagger} \hat{a} + \hat{b} \hat{b}^{\dagger})/2$, $\hat{a}\,\hat{b}$ and $\hat{a}^{\dagger} \,\hat{b}^{\dagger}$ (all entering the total Hamiltonian) satisfy the commutation relations of the $SU(1,1)$ Lie algebra. From this observation Eq. (\ref{EQP}) can be related to the Wigner matrix elements of the unitary and irreducible representations of the corresponding group; the states $|n_{a}\, n_{b}\rangle$ (where $n_{a}$ and $n_{b}$ denote, respectively, the eigenvalues of the number operators of the system and of the environment) form a basis for the irreducible representations;  $k$ (the so-called Bargmann parameter \cite{BAR} related to the eigenvalue of the Casimir operator) corresponds to $k = (1+Q)/2$ where $Q$ is now the eigenvalue of $\hat{Q} =\hat{N}_{a} - \hat{N}_{b}$; note, in fact, that $[\hat{H}, \hat{Q}] =0$.
\vspace{-0.4cm}
\begin{figure}[ht!]
\begin{centering}
\includegraphics[width=4.2cm,height=4.5cm]{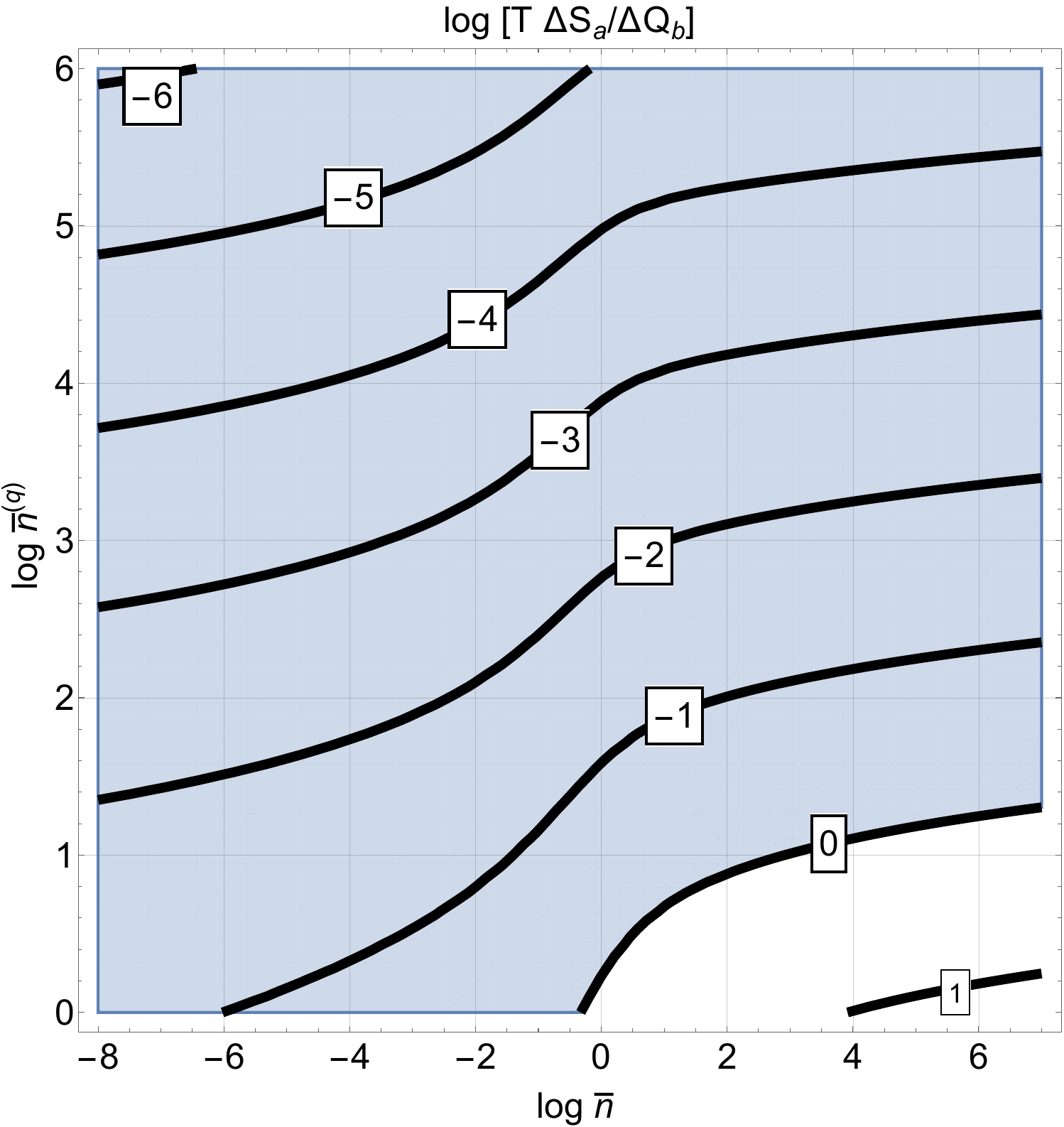}
\includegraphics[width=4.2cm,height=4.5cm]{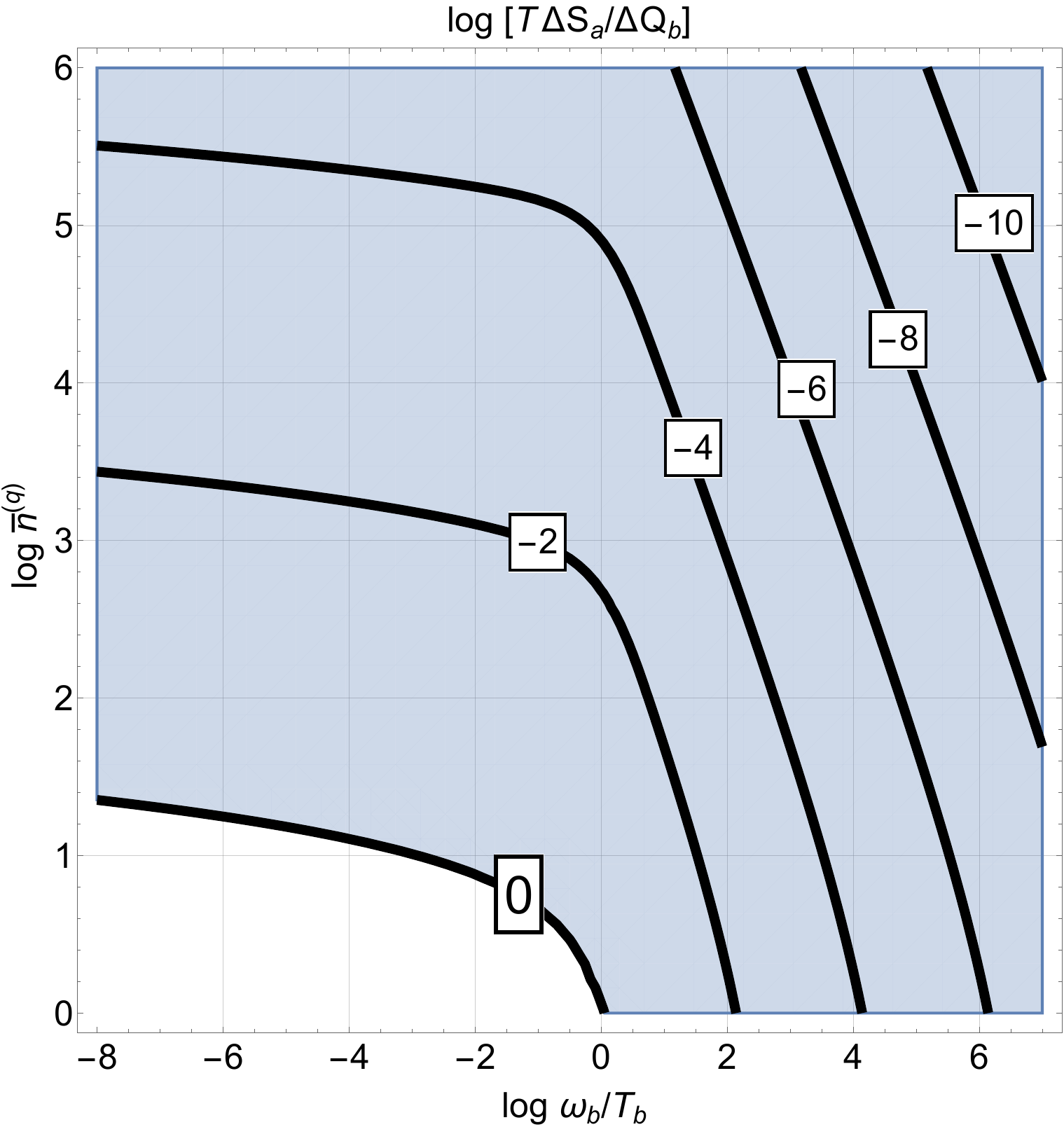}
\par\end{centering}
\caption{\label{FIGU1} The values reported in the labels are constant along each contour and correspond to the common logarithms of $T  \Delta S_{a}/\Delta Q_{b}$ computed from the right-hand side of Eq. (\ref{EQR}). In the shaded regions the bound of Eq. (\ref{EQA1}) is satisfied. The plane ($\overline{n}$, $\overline{n}^{(q)}$) is illustrated in the left plot; in the right plot we traded instead  $\overline{n}$ for $\omega_{b}/T$. Common logarithms are employed on both axes of each plot. The areas where the bound is not satisfied, as anticipated in Eq. (\ref{EQA1}), correspond to large thermal multiplicities and comparatively negligible quantum amplification (i.e. $\overline{n}^{(q)} = {\mathcal O}(1)$ or smaller). }
\vspace{-0.2cm}
\end{figure}

The result of Eq. (\ref{EQA1}) follows from Eqs. (\ref{EQO})--(\ref{EQP}) since, after computing the entropy variation of the system (i.e. $\Delta\, S_{a}= S[ \hat{\rho}_{a}^{(fin)}] - S[ \hat{\rho}_{a}^{(in)}]$), we directly obtain $\Delta S_{a}= [(\overline{N} +1) \ln{(\overline{N}+1)} - \overline{N} \ln{\overline{N}}]$
where $\overline{N} = \overline{n}^{(q)}(\overline{n} +1)$ now indicates, with stenographic notation, the total averaged multiplicity of the final state. Moreover the heat flowing to the environment can be also expressed in terms of $\overline{N}$, i.e. $\Delta Q_{b} = \mathrm{Tr}[ \hat{H}_{b}( \hat{\rho}_{b}^{(fin)} -  \hat{\rho}_{b}^{(in)})]= \omega_{b} \, \overline{N}$; thus the ratio between $T \Delta S_{a}$ and $\Delta Q_{b}$ finally gives:
\vspace{-0.1cm}
\begin{equation}
T \frac{\Delta S_{a}}{\Delta Q_{b}}= \biggl(\frac{T}{\omega_{b}}\biggr)\biggl[\frac{\ln{\overline{N}}}{\overline{N}} + \biggl(1 + \frac{1}{\overline{N}}\biggr) \ln{\biggl(1 + \frac{1}{\overline{N}}\biggr)}\biggr].
\label{EQR}
\vspace{-0.1cm}
\end{equation}
Because $\omega_{b}/T$ can be traded for the averaged thermal multiplicity $\overline{n} = (e^{\omega_{b}/T} -1)^{-1}$, the right-hand side of Eq. (\ref{EQR}) only depends on $\overline{n}$ and $\overline{n}^{(q)}$. Furthermore, despite the value of $\overline{n}$, the value of $\overline{N}$ always exceed $1$ provided the quantum amplification is effective (i.e.  $\overline{n}^{(q)} \geq 1$). All in all, when $\omega_{b} \geq T$ and $\overline{N} > 1$ the bound of Eq. (\ref{EQA1}) is verified since, in this limit, the right-hand side of Eq. (\ref{EQR}) never overshoots $1$.
In the complementary region of the parameter space (i.e. $\omega_{b}< T$),  $\overline{n} \simeq (T/\omega_{b})>1$ so that the condition stemming from Eqs. (\ref{EQA1}) and (\ref{EQR}) reads $T \Delta S_{a}/\Delta Q_{b} = [1+ \ln{(\overline{n}^{(q)}\, \overline{n})}]/\overline{n}^{(q)} <1$; this requirement is satisfied provided $\overline{n} \leq e^{[\overline{n}^{(q)}-1]}/\overline{n}^{(q)}$, as anticipated after Eq. (\ref{EQA1}). In Fig. \ref{FIGU1} the logarithms of the right-hand side of Eq. (\ref{EQR}) are illustrated without approximations: the plane $(\overline{n},\, \overline{n}^{(q)})$ is examined in the left panel while in the right plot the plane $(\omega_{b}/T, \,\overline{n}^{(q)})$ is analyzed. The shaded areas correspond to the regions where the bound of Eq. (\ref{EQA1}) is satisfied and $T\,  \Delta S_{a}/\Delta Q_{b}<1$.

The production of particles in curved backgrounds ultimately rests on the same physical premises of the quantum parametric amplification \cite{PAR,FORD,GRSID,PARTO} and this is why pairs of gravitons are  produced thanks to the early variation of the space-time curvature \cite{GR1,GR2}. To substantiate this statement we now introduce the second-order tensor fluctuation of the Einstein-Hilbert action in a spatially flat Friedmann-Robertson-Walker background \cite{FP} (see also \cite{MAC}) 
\vspace{-0.2cm}
\begin{equation}
S_{g} = \frac{1}{8\ell_{P}^2} \int \, d^{4} x \,a^2(\tau) \, \eta^{\mu\nu}\, \partial_{\mu} \, h_{i\, j} \, \partial_{\nu} h^{i\, j},
\vspace{-0.2cm}
\label{EQS}
\end{equation} 
where $h_{i\,j}$ is solenoidal, traceless (i.e. $h_{i}^{\,\, i} = \partial_{i} h^{i\, j} =0$) and it describes the tensor modes of the four-dimensional geometry. In Eq. (\ref{EQS}) $\eta_{\mu\nu}$ denotes the Minkowski metric [with signature $(+, -, -, -)$] and $a(\tau)$ is the scale factor, written as a function of the conformal time coordinate $\tau$. The relative variation of the scale factor is given by ${\mathcal H}= (\ln{a})^{\prime}$ and the prime denotes hereunder, for the sake of conciseness, a derivation with respect to $\tau$. When the rescaled canonical amplitudes $\mu_{i\, j}= h_{i\,j}/a(\tau)$ and the comoving momenta $\pi_{i\,j} = (\partial_{\tau} \mu_{i\, j} - {\mathcal H} \mu_{i\, j})/(8\ell_{P}^2)$ are promoted to the status of field operators we deduce 
\begin{eqnarray}
\hspace{-0.7cm}
&&\hat{\mu}_{i\,j}(\vec{x},\tau) = \sqrt{2} \, \ell_{P}\, \int \frac{d^{3} k}{ (2 \pi)^{3/2}}\sum_{\alpha}\, e_{i\, j}^{(\alpha)} \hat{\mu}_{\vec{k}, \, \alpha} e^{- i \vec{k}\cdot\vec{x}}, 
\nonumber\\
\hspace{-0.7cm}
&&\hat{\pi}_{i\,j}(\vec{x},\tau) = \frac{1}{4 \sqrt{2} \ell_{P}}\int  \frac{d^{3} k}{ (2 \pi)^{3/2}}\sum_{\alpha}\, e_{i\, j}^{(\alpha)}\hat{\pi}_{\vec{k}, \, \alpha} e^{- i \vec{k}\cdot\vec{x}},
\label{EQT}
\end{eqnarray}
where the sums run over the two tensor polarizations $\alpha = \oplus, \otimes$, i.e.
$e_{i\,j}^{(\oplus)}(\hat{k}) = (\hat{m}_{i} \, \hat{m}_{j} + \hat{n}_{i} \, \hat{n}_{j})$ and  
$e^{(\otimes)}_{i\,j}(\hat{k}) = (\hat{m}_{i} \, \hat{n}_{j} - \hat{m}_{j} \, \hat{n}_{i})$; $\hat{m}$, $\hat{n}$ and $\hat{k}$ are just a triplet of mutually orthogonal unit vectors obeying $\hat{m} \times \hat{n}= \hat{k}$.

In terms of the creation and annihilation operators with opposite three-momenta we have $\hat{\mu}_{\vec{k}, \, \alpha} = ( \hat{a}_{\vec{k},\, \alpha} + \hat{a}_{-\vec{k},\, \alpha}^{\dagger})/\sqrt{2 \, k}$ and $ \hat{\pi}_{\vec{k}, \, \alpha} = - i\, ( \hat{a}_{\vec{k},\, \alpha} - \hat{a}_{-\vec{k},\, \alpha}^{\dagger})\, \sqrt{k/2}$. The Hamiltonian operator deduced from the action (\ref{EQS}) takes then the same form of Eqs. (\ref{EQB}) and (\ref{EQG}), 
 \vspace{-0.2cm}
\begin{eqnarray}
\hspace{-0.6cm}
\hat{H}_{g}(\tau) &=& \frac{1}{2} \int d^{3}k \sum_{\alpha} \, k\, \bigl[ \hat{a}_{\vec{k},\,\alpha}^{\dagger} \hat{a}_{\vec{k},\, \alpha} + 
\hat{a}_{-\vec{k},\, \alpha}\hat{a}_{-\vec{k},\, \alpha}^{\dagger}\bigr]
\nonumber\\
\hspace{-0.6cm}
&+&\frac{1}{2} \int d^{3}k \sum_{\alpha} \,\bigl[ \lambda^{\ast} \hat{a}_{\vec{k},\, \alpha} \hat{a}_{-\vec{k},\, \alpha} + \lambda
\hat{a}_{\vec{k},\, \alpha}^{\dagger} \hat{a}_{-\vec{k},\, \alpha}^{\dagger} \bigr],
\label{EQU}
\vspace{-0.5cm}
\end{eqnarray}
where $\lambda= i {\mathcal H}$. The previous quantum mechanical analysis can be now be repeated by bearing in mind that the modes of the field with opposite three-momenta now operate in different subspaces of the Hilbert space. For instance the analog of Eq. (\ref{EQI}) can now be written in terms of the complex functions $u_{p,\,\alpha}(\tau,\tau_{in})$ and $v_{p,\,\alpha}(\tau,\tau_{in})$ subjected to the unitarity condition $|u_{p,\,\alpha}^{\ast}(\tau,\tau_{in})|^2 - |v_{p,\,\alpha}^{\ast}(\tau,\tau_{in})|^2 =1$
\vspace{-0.2cm}
\begin{eqnarray}
\hspace{-0.5cm}
&& \hat{a}_{\vec{p},\,\alpha}(\tau) = u_{p,\,\alpha}(\tau,\tau_{in}) 
\hat{b}_{\vec{p},\,\alpha}^{(in)} -  
v_{p,\,\alpha}(\tau,\tau_{in}) \hat{b}_{-\vec{p},\,\alpha}^{(in)\,\dagger},  
\nonumber\\
\hspace{-0.5cm}
&& \hat{a}_{- \vec{p},\alpha}^{\dagger}(\tau) = u_{p,\,\alpha}^{\ast}(\tau,\tau_{in}) \hat{b}_{-\vec{p},\,\alpha}^{(in)\dagger} -  
v_{p,\,\alpha}^{\ast}(\tau,\tau_{in}) \hat{b}_{\vec{p},\,\alpha}^{(in)}.
\label{EQV}
\vspace{-0.5cm}
\end{eqnarray}
The two functions of Eq. (\ref{EQV}) are however determined by the following pair of dynamical equations 
\vspace{-0.2cm}
\begin{equation}
u_{p,\alpha}^{\prime} = - i p u_{p,\alpha} - {\mathcal H} v_{p,\alpha}^{\ast},\quad
v_{p,\alpha}^{\prime} = - i p v_{p,\alpha} - {\mathcal H} u_{p,\alpha}^{\ast},
\label{EQZ}
\end{equation}
that must be solved with the appropriate boundary conditions. The 
averaged multiplicity at $\tau_{fin}$ becomes now $\overline{n}_{k}^{(q)} =  |v_{p,\,\alpha}^{\ast}(\tau_{fin},\tau_{in})|^2 $ and it depends, as the initial thermal multiplicity $\overline{n}_{k} = ( e^{k/T} -1)^{-1}$, on the modulus of the comoving three-momentum $k$. The quantum states  are classified in the Fock basis where the oscillators $-\vec{k}$ correspond to the system while the ones with $+\vec{k}$ are associated with the environment. The entropy increase for each $k$-mode has the same form of the one already discussed in Eq. (\ref{EQR}) with the difference that, in the present context, all the average multiplicities depend on $k= |\vec{k}|$, i.e. $\overline{N}_{k} = \overline{n}^{(q)}_{k} [ \overline{n}_{k} +1]$. The bound of Eq. (\ref{EQA1}) holds then for each mode of the field, i.e. $T  \Delta S_{k}/\Delta Q_{k} <1$. The condition $\overline{n}^{(q)}_{k} \gg  1$ is naturally satisfied since the produced number of graviton pairs exceeds $1$ except for the maximal wavenumber (or maximal frequency) of the spectrum.
\vspace{-0.2cm}
\begin{figure}[ht!]
\begin{centering}
\includegraphics[width=4.2cm,height=4.5cm]{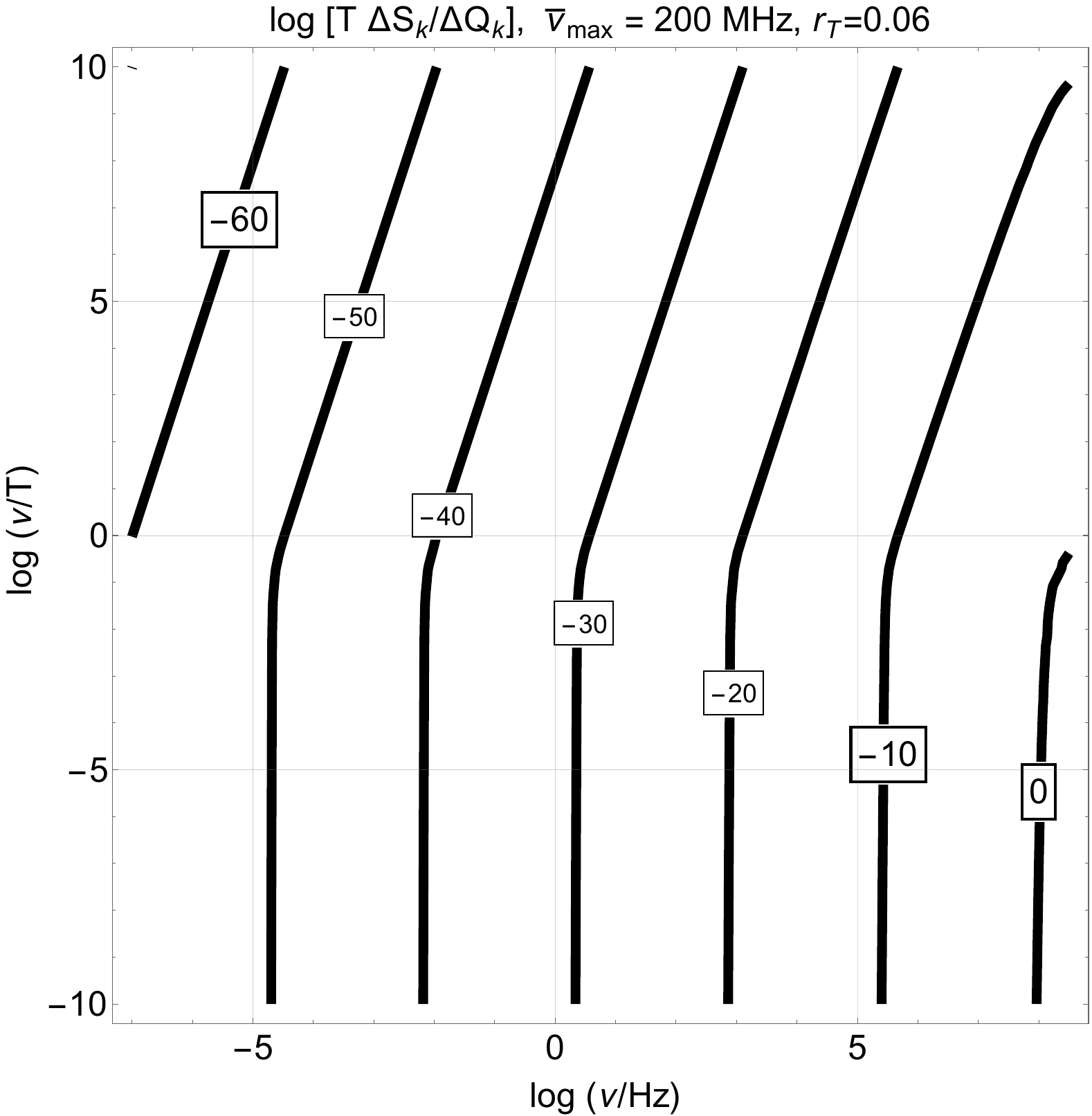}
\includegraphics[width=4.2cm,height=4.5cm]{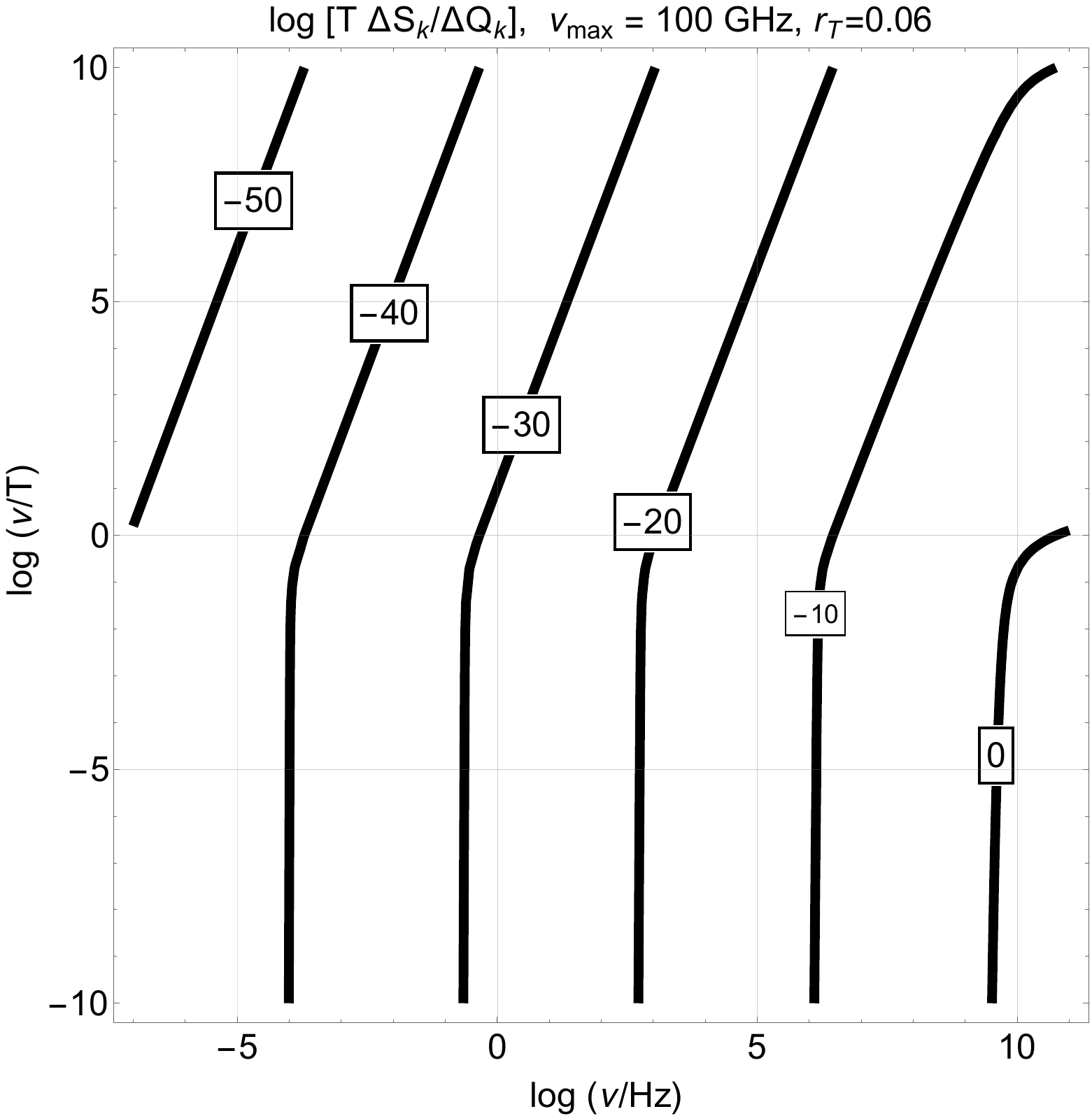}
\par\end{centering}
\caption{\label{FIGU2} As in Fig. \ref{FIGU1}, the different labels in both plots indicate the logarithms of $T  \Delta S_{k}/\Delta Q_{k}$. In the left panel the 
bound is illustrated in the context of the concordance paradigm (i.e. $\overline{\nu}_{max} = 200 \, \mathrm{MHz}$, $\delta =1$). In the right plot we consider instead the example $\nu_{max} = 100$ GHz and
$\delta =1/2$. When $\nu= {\mathcal O}(\nu_{max})$ we have that $\overline{n}_{k}^{(q)} = {\mathcal O}(1)$ and a single pair of gravitons is produced for each mode of the field. Common logarithms are employed 
on both axes.}
\vspace{-0.2cm}
\end{figure}

It is actually known that in terms of the comoving frequency (i.e. $\nu = 2 \pi k$) the spectrum of the relic gravitons extends between the aHz region ($\mathrm{aHz} = 10^{-18}\, \mathrm{Hz}$) and the THz domain (see, for instance, \cite{rr1}). For a broad range of scenarios the frequency dependence of the averaged multiplicity can be expressed as $\overline{n}^{(q)}(\nu,\tau_{0}) = (\nu/\nu_{max})^{- 4 + m_{T}}$ \cite{MG2} where $m_{T}$ indicates the spectral index; the value of the maximal frequency corresponds to the production of a single graviton pair. Both
$\nu_{max}$ and $m_{T}$ are model-dependent even though $\nu_{max}$ cannot exceed the THz \cite{MG3}. For a single post-inflationary stage of expansion prior to matter-radiation equality $m_{T} = [32 (1 - \delta)+ 2 r_{T}(\delta -2)]/(16 - r_{T})$ where $r_{T}< 0.06$ is the tensor-to scalar ratio (bounded by the observations of the temperature and polarization anisotropies of the Cosmic Microwave Background \cite{RR1}) while $\delta$ approximately accounts for the rate of post-inflationary expansion.
In the concordance paradigm (i.e. $\delta \to 1$, $m_{T} = - r_{T}/8$) we have $\overline{\nu}_{max} = {\mathcal O}(200) \, \mathrm{MHz}$; this value can increase up to ${\mathcal O}(100)$ GHz if the post-inflationary stage of expansion is different from radiation (i.e. $\delta \neq 1$) and $\nu_{max}$ can be either larger or smaller than $\overline{\nu}_{max}$. For instance the averaged multiplicity is of the order of $10^{20}$ in the concordance scenario
 for typical frequencies ${\mathcal O}(\mathrm{kHz})$ (corresponding to the region of operating wide-band detectors 
 \cite{MG2}).

In Fig. \ref{FIGU2} we illustrate the bound of Eq. (\ref{EQA1}) for frequencies ranging between the aHz and the THz; the two sets of parameter correspond, respectively, to the concordance scenario and to the presence of a post-inflationary phase expanding slower than radiation (with $\delta =1/2$). In more general terms, if we demand that the bound is satisfied without any constraint on the initial thermal multiplicity, then  
$(T/\nu)< 1$ where, as before, $T$ is the pre-inflationary temperature of the thermal gravitons defining the initial environment. In terms of the present frequencies  $(\nu/T)$ depends then on the duration of inflation measured by ${\mathcal N}$ (i.e. the total number of $e$-folds):
\vspace{-0.2cm}
\begin{equation}
\nu/T= 1.3\times 10^{-2}(\nu/\nu_{p})\,(T_{max}/T) \,\exp{[ {\mathcal N} - {\mathcal N}_{c}]},
\label{EQY}
\vspace{-0.1cm}
\end{equation}
where $\nu_{p}= {\mathcal O}(3) \mathrm{aHz}$ is the lowest frequency of the gravitons \cite{GR1,GR2,FP,AA7,AA8,AA9,rr1}; $T< T_{max} = [45/(4 \pi^3 g_{eff})]^{1/4} \, \sqrt{H_{max} M_{P}}$ is the maximal temperature 
of the thermal gravitons compatible with early occurrence of inflation and $H_{max}\leq 10^{-5} M_{P}$ denotes the expansion rate at the onset of the inflationary stage; finally $g_{eff}$ is the effective number of relativistic degrees of freedom at $T_{max}$ ($g_{eff} =106.75$ in the case of a standard particle content \cite{weinberg}). Since $ {\mathcal N}_{c} = {\mathcal O}(60)$ is the minimal number of $e$-folds required to solve the problems associated with the causal structure of the hot big-bang model \cite{weinberg}, the stronger version of the entropy bound obtained here demands $\nu/T >1$. This either implies that 
${\mathcal N} = {\mathcal O}({\mathcal N}_{c})$ and $ T \leq 10^{-2} T_{max}$ or ${\mathcal N} > {\mathcal N}_{c}$ in Eq. (\ref{EQY}) for all frequencies of the spectrum (i.e. $\nu > \mathrm{aHz}$). We remark that a total number of $e$-folds larger than ${\mathcal O}(60)$ is generally required for inflationary scenarios without fine-tuning.

{\it Conclusions.} To delete information the entropy of the system must be reduced but this decrement demands an energy cost. When the entropy variation of the system is positive the Landauer's bound is however not constraining. From a microscopic viewpoint we derived a bound that limits the increase of the Von Neumann entropy associated with the quantum parametric amplification. We showed, as an application, that if the obtained limit is enforced over the whole spectrum of the relic gravitons the total number of inflationary $e$-folds must roughly exceed the critical number ${\mathcal O}(60)$ required to address the causality problems of the hot big-bang scenario.

It is a pleasure to thank the kind assistance of A. Gentil-Beccot, P. Birtwistle, A. Kohls, L. Pieper, S. Rohr and J. Vigen and of the whole CERN Scientific information Service.

\end{document}